\documentclass[12pt,preprint]{aastex}

\slugcomment{Accepted by \it The Astronomical Journal \rm}
\shorttitle{Baryons and Dark Matter in Halos}
\shortauthors{Putman et al.}
\received{2005 July 13}

\def\gtrapprox{\;\lower 0.5ex\hbox{$\buildrel >\over \sim\ $}}
\def\lessapprox{\;\lower 0.5ex\hbox{$\buildrel < \over \sim\ $}}

\def\Msun  {${\rm M}_\odot$}
\def\deg   {$^\circ$}

\def\kms   {\ km s$^{-1}$}

\def\lya   {Ly$\alpha$ }
\def\Lya   {Ly$\alpha$ }

\begin{document}
\title{The Relationship Between Baryons and Dark Matter in Extended Galaxy Halos}
\author{M. E. Putman\altaffilmark{1}, J. L. Rosenberg\altaffilmark{2}, J. T. Stocke\altaffilmark{3}, R. McEntaffer\altaffilmark{3}}

\altaffiltext{1}{Department of Astronomy, University of Michigan, 830 Dennison, 500 Church St, Ann Arbor, MI 48109; mputman@umich.edu}
\altaffiltext{2}{Harvard-Smithsonian Center for Astrophysics, 60 Garden Street MS 65,
Cambridge, MA 02138; jlrosenberg@cfa.harvard.edu}
\altaffiltext{3}{Center for Astrophysics and Space Astronomy, University of Colorado, Boulder, CO 80309-0389; stocke@colorado.edu} 

\newpage

\begin{abstract} 
The relationship between gas-rich galaxies and \lya absorbers is addressed in this paper in the context of the baryonic
content of galaxy halos.
Deep Arecibo HI observations are presented of two gas-rich spiral galaxies within $125~h^{-1}_{70}$ kpc projected distance of a \Lya absorber 
at a similar velocity.  The galaxies investigated are close to edge-on and the absorbers lie almost
along their major axes, allowing for a comparison of the \lya absorber velocities with galactic rotation. 
This comparison is used to examine whether the
absorbers are diffuse gas rotating with the galaxies' halos, outflow material from
the galaxies, or intergalactic gas in the low redshift cosmic web. The results
indicate that if the gas resides in the galaxies' halos it is not rotating with
the system and possibly counter-rotating. In
addition, simple geometry indicates the gas was not ejected from
the galaxies and
there are no gas-rich satellites detected down to $3.6 - 7.5 \times 10^{6}$ \Msun, or remnants of 
satellites to $5-6 \times 10^{18}$ cm$^{-2}$.   The gas could potentially be infalling from
large radii, but the velocities and distances are rather high compared to the 
high velocity clouds around the Milky Way.  The most likely explanation is the galaxies and absorbers
are not directly associated, despite the vicinity of
the spiral galaxies to the absorbers ($58-77~h^{-1}_{70}$ kpc from the HI edge).
The spiral galaxies reside in a filament of intergalactic gas, and the gas detected by the absorber 
has not yet come into equilibrium with the galaxy.  These results also indicate that the massive, extended dark
matter halos of spiral galaxies do not commonly have an associated diffuse baryonic component at large radii.
\end{abstract}

\keywords{galaxies: kinematics and dynamics $-$ galaxies: 
spiral $-$ galaxies: halo $-$ intergalactic medium $-$ dark matter $-$ large scale structure of the universe}

\section{Introduction}

Gaseous envelopes surrounding galaxies out to radii of 100-200 kpc are
a common prediction of galaxy formation models (e.g., White \& Rees 1978; Maller \& Bullock 2004).
Generally the dark matter and gas in galaxy halos are assumed to be coupled.
As the gas falls onto the galaxy it comes into equilibrium with the dark halo and
is subsequently able to cool and rain down onto the galaxy's disk.  Absorption line
experiments are one of the sole methods of tracing the low density gas as it cools
and falls onto the galaxy.  Therefore, comparing the kinematics of a galaxy's dark matter
halo to absorption line systems in the halo is a direct test of the radius at which
the gas comes into equilibrium with the dark matter halo and the dark matter$-$baryon connection.

Several studies have suggested that a large fraction of the Ly$\alpha$ absorbers with column densities of 
N$_{HI} \sim 10^{13-17.3}$ cm$^{-2}$ reside in galaxy halos (e.g., Chen et al. 2001; Bowen, Pettini, \& Blades 2002). 
This has generally been derived from a correspondence between the velocity of the absorber and systemic velocity of the galaxy. 
Chen et al. examined 34 galaxy - absorber pairs over a wide range of
redshifts ($z$ = 0.075 - 0.892), impact parameters (18 - 250 $h^{-1}_{70}$ kpc), and galaxy luminosities 
(low surface brightness - $\sim3$L$_*$) to determine how the extent
of a galaxy's diffuse gaseous halo scales with galaxy properties.
They concluded that the typical L$_*$ galaxy is surrounded
by an extended gaseous halo of radius $\sim 260~h^{-1}_{70}$ kpc with a covering fraction of 94\% for
gas with N$_{\rm HI} > 10^{14}$ cm$^{-2}$.  Bowen et al. (2002) studied a sample of 8 galaxies at 
small projected distances from a QSO and found that within 285 $h^{-1}_{70}$ kpc of a galaxy their 
results implied a covering factor 
of $\sim 100$\% for gas with N$_{\rm HI} > 10^{13}$ cm$^{-2}$.
These results suggest the galaxy formation scenario outlined above
is on the right track for reproducing the observed baryon content of extended galaxy halos.

It has also been suggested that these low column density \lya absorbers ($\sim 10^{14}$ cm$^{-2}$) 
are not related to individual galaxies,
but are part of the extended cosmic web (e.g., Dav\'e et al. 1999; Penton, Stocke \& Shull 2002; Stocke et al. 2005).
In various CDM simulations the \lya absorbers trace fluctuations in the baryon density originating from the epoch of
structure formation (e.g., Dav\'e et al. 1999; Zhang et al. 1998).  Dav\'e et al. find that at every redshift these low
column density \lya absorbers 
originate from diffuse gas in the unshocked intergalactic medium (IGM), while stronger absorbers are
generally found close to galaxies.  This is observationally supported by the work of Penton et al. (2002), who find
the median distance between a low column density \lya absorber and a galaxy is $\sim 500~h^{-1}_{70}$ kpc, 
twice the median distance between bright galaxies in their sample.  They therefore suggest that the majority of the low redshift \lya absorbers
are associated with large-scale structures, rather than individual galaxies.

To further investigate the relationship between galaxies and \lya absorbers, studies
correlating absorption line velocities with the detailed kinematics of the associated galaxy
have recently been completed (Steidel et al. 2002; Keeney et al. 2005; Cot\'e et al. 2005, hereafter C05).  
Steidel et al. examined Mg II absorption line systems, or gas with column densities N$_{HI} > 10^{17}$ cm$^{-2}$,
at redshifts of 0.44 $\le z \le 0.66$. 
In the simulations discussed above, these strong absorbers are generally associated with the gaseous halos 
surrounding L$_*$ spiral galaxies.  They
found in all 5 of their cases the velocity offset of the absorption line was in the right
sense to be qualitatively explained by an extension of the disk rotation to the line of 
sight.  (The lines of sight are well beyond where the optical rotation curves were measured, by a factor of 2 to 
as much as a factor of 6.)
The extension of disk rotation was not a simple rotation model however; the absorption line
velocities had to be explained with thick rotating halo gas layers (Charlton \& Churchill 1998).
In contrast to the expected results of gas at large radii following the overall kinematics
of the galaxy, the work of C05 finds several examples of gas which is not in
rotation with the galaxy.  C05 examined low column density absorption line
systems (N$_{\rm HI} < 10^{14}$ cm$^{-2}$) between $59 - 181~h_{70}^{-1}$ kpc from the
center of a spiral galaxy.
In their sample of 5 low column density absorption line systems near galaxies, the velocities of two absorbers can be fit
by an extended warp or thick rotating gas layers.
The other three systems can not be fit by any sort of extended rotating disk.  In addition, Ly$\alpha$ absorption was 
not detected along 3 lines of sight that were more than $265~h^{-1}_{70}$ kpc from the neighboring spiral galaxy.

Historically, the link between \lya absorbers and galaxies has been investigated using optical
searches for galaxies in the vicinity of the absorbers (e.g., McLin et al. 2002; 
Penton et al. 2002; Impey, Petry, \& Flint 1999). 
Since the absorbers trace the warm-hot {\it gaseous} intergalactic medium (IGM), an
obvious comparison is to study the {\it gaseous} components of galaxies.
In addition, comparing the detailed kinematics of the galaxies to the absorption line systems 
has been limited to the studies discussed above, with inconclusive results as to the
actual relationship.
Here we present Arecibo\footnote{The Arecibo Observatory is part of the National Astronomy and 
Ionosphere Center, which is operated by Cornell University under a cooperative agreement with the National 
Science Foundation.} observations of two edge-on spiral galaxies
with \lya absorbers $< 125~h^{-1}_{70}$ kpc along the major axis of the galaxies. 
Several questions with regard to the relationship between \lya absorbers and
galaxies are addressed, including:   Do the \lya absorbers near galaxies trace gas in the galaxies' extended dark matter halos, and
if so, is the gas rotating with the dark 
matter halos?; Do the \lya absorbers near galaxies
represent galactic waste, galactic fuel, or are they simply part of large scale cosmic filaments?;
Do the \lya absorbers represent the presence of a low surface brightness galaxy, rich in gas, but not
easily detected in the optical?  This paper proceeds by describing the Arecibo observations and data reduction, and
subsequently presents the results of the observations.  Finally, the implications of the results are
discussed in the context of the questions presented above.  

\section{Observations and Data Reduction}

This paper presents neutral hydrogen data of the regions around 2 \lya absorbers in the 
sightlines towards PG 1211+143 and Ton 1542 that are at similar velocities to the nearby spiral 
galaxies IC 3061 and UGC 7697, respectively.  
These HI data are from the Arecibo Radio Telescope and were observed as part of a
large project to map the gaseous environment of 17 low-redshift Ly$\alpha$ absorbers along 4 sightlines.  
Driftscan maps 0.5\deg $\times$ 0.5\deg~in size were made in November and 
December 2001 and April to May of 2002 around the 4 sightlines.  Each individual drift 
scan was spaced by 50\arcsec, and the integration time per beam was 14.4 seconds.  Twenty-one driftscan
maps were observed for each source, resulting in a total integration time per beam of
302.4 seconds.
A calibration diode (cal) was fired at the beginning and end of each scan.  Cals that were fired
while there was a continuum source in the beam were excluded from the calibration, as they do
not give an accurate measurement of the system response.
The beam size in the final gridded cubes is 215\arcsec, or $\sim3.6$\arcmin, and the velocity coverage extends 
from approximately $-$1035 \kms~ to 12,750 \kms.  For this paper we concentrate on the cubes
centered on the background sources PG 1211+143 and Ton 1542 as these were the only cubes which
had absorbers at a similar velocity to a gas-rich spiral galaxy.   
The other two sightlines and low redshift \lya absorbers will be discussed in a future paper.
The sensitivity achieved in each 5 \kms~channel
is $\sim2$ mJy/bm or 26 mK (1$\sigma$) for the cube centered on PG 1211+143 and 
$\sim1.7$ mJy/bm or 22 mK (1$\sigma$) for
the cube centered on Ton 1542.   These values were calculated with the MIRIAD imstat routine on line-free channels
in the vicinity of the galaxies of interest (IC 3061 and UGC 7697, respectively).    These limits correspond to a 5$\sigma$ HI mass limit of 
5200 D$^{2}$(Mpc) and 4420  D$^{2}$(Mpc) ($\Delta$v = 50 \kms), respectively.  Using the distances listed in Table 1 for
the galaxies of interest here, this
corresponds to 7.5 $\times 10^{6}$ \Msun~at 38 Mpc for the cube centered on PG 1211+143 and 3.6 $\times 10^{6}$ \Msun~at 28.7 Mpc for 
the cube centered on Ton 1542.

The data reduction was completed with IDL routines developed at the observatory.  A robust median fit 
was originally used for the bandpass correction, but was found not to work in the strips that
contained both a large galaxy and continuum sources because the median value contained emission.  
The solution that resulted in the maps presented here is similar to the method presented for
recovering extended sources in the HIPASS data (Putman et al. 2003a).  Each strip
of 120 samples was broken up into 3 sections, the median was computed for each section, and
the minimum of these medians was used for the final bandpass correction.
The system temperature was subsequently removed using a robust mean over the channels of the spectra.
Twenty-one 0.5\deg $\times$ 0.5\deg~ maps 
(composed of 36 scans each)  were subsequently averaged together to obtain the HI
data presented here.  In the final calculation of the galaxies' HI masses (Table 1) and presentation of
the spectra (Figures 4 and 8), a second order polynomial was fit to a $\sim 500$ \kms~
range around each spectrum.

The absorption line data for the absorber closest in velocity to the large spiral
galaxy for PG 1211+143 and Ton 1542 are summarized in Table 2.  These
absorbers are taken from the sample of Penton et al. (2004) where the STIS spectra and details
of the observation and reduction are
presented.  Both absorption line profiles appear to be single components. For PG 1211+143 we have also 
looked at the archived STIS E140M spectrum to confirm that the line appears to be a single component at higher resolution. 

\section{Results}
\subsection{IC 3061 and the PG 1211+143 Sightline}

Figure 1 shows the HI distribution of IC 3061 as observed by Arecibo over a $25^{\prime} \times 25^{\prime}$ 
Digital Sky Survey image centered on the galaxy.  
The entire field of our Arecibo observations is shown in velocity map of Figure 2, as well as the beam in the lower right corner. 
The position of the background source (PG 1211+143) used to detect the absorption line is shown in both plots.
Unfortunately, IC 3061 is at the edge of the field so a small fraction of the galaxy
has been cut-off.  The data presented have a uniform number of drift scans across the map.
IC 3061 is the only galaxy detected in the data cube from $-1200 - 5000$ \kms.  There are also no other 
galaxies present in this spatial and velocity region of the sky as catalogued by NED (considering only those
galaxies with measured redshifts).
The first galaxy to appear in the cube after IC 3061 is at $\sim 6600$ \kms, nicely corresponding to the
velocity of an absorber (6623 \kms; Penton et al. 2002).  This and other objects detected in this cube will
be discussed in Rosenberg et al. (2005).  
The galaxy at the lower left of Fig. 1 is NGC 4208 at -81 \kms~ (de Vaucouleurs et al. 1991), and was
off the edge of our HI survey. 

Table 1 lists the properties of IC 3061 from this survey and NED.
The HI mass of this galaxy is $5.3 \times 10^{9}$ \Msun~from our Arecibo data, using a distance
of 38.0 Mpc from the Tully-Fisher relation (Solanes et al. 2002).  A previous calculation of this galaxy's HI mass with
Arecibo, using integrated 21-cm line profiles and applying a correction for the galaxies extended nature, 
is $4.3 \times 10^{9}$ \Msun~(Haynes \& Giovanelli 1986).
The dynamical mass of IC 3061 is $2.8 \times 10^{11}$ \Msun, and was calculated using M$_{\rm dyn}$ $=$ r$_{\rm HI}$V$^{2}/$G and
V$= \Delta$v$/2$sin$i$, where r$_{\rm HI} = 52~h^{-1}_{70}$ kpc (these data), $\Delta$v$ =310$\kms~(these data) and $i= $82\deg~(Huchtmeier \& Richter 1989). 
The sightline of the absorber is 11\arcmin~from the optical center of IC 3061 
and 7\arcmin~ from the edge of the observed rotating HI disk.  In projection, at a distance of 38 Mpc, this is
$122~h^{-1}_{70}$ kpc from the optical center and $77~h^{-1}_{70}$ kpc from the
1.4 $\times 10^{19}$ cm$^{-2}$ column density contour at the HI edge.   Note that our estimate
for the size of the HI disk is severely limited by our beamsize which is $40~h^{-1}_{70}$ kpc 
at the distance of the galaxy.

Figure 2 depicts the velocity distribution of IC 3061 with the contours of the integrated intensity map overlaid and
the position of the absorber sightline labeled.  This figure represents the entire field of view observed
by our Arecibo map.  The velocity of the galaxy extends from $\sim$2150 \kms~to $\sim$2490 \kms.  
The absorber along the PG 1211+143 sightline with the closest velocity to this galaxy is at 2130 \kms, lower 
than the central velocity of the galaxy despite being on the receding side of the disk. 
The properties of this absorption line are listed in Table 2.  
The absorber is indistinguishable from a single line profile
and has a column density of 10$^{13.76}$ cm$^{-2}$, 
while the column density limit of our HI observations per channel (5 \kms)
is $1.2 \times 10^{18}$ cm$^{-2}$ (5$\sigma$).   
The closest absorber in velocity to
the 2130 \kms~absorber is at 4944 \kms (Penton et al. 2004).
The projected position of
the absorber is 20\deg~from being along the major axis of the galaxy as shown in Figures 1 and 2.
A rotation curve was not fit to this data due to the limited number of independent beams across the galaxy.
 Figure 3 depicts a right ascension-velocity map through IC 3061 at the declination of the absorber.
This figure also shows that the velocity of the absorber is closer to the velocities on the far side of the galaxy 
disk rather than on the near side.  Table 3 outlines the spatial and kinematic differences between
IC 3061 and the absorber.

Figure 4 shows the integrated HI spectrum of IC 3061.   It is a beautiful double-horned profile,
representative of its galaxy type (SBc) and edge-on nature.  
IC 3061 is included in the Flat Galaxy Catalog of Karachentsev et al. (1993) and is considered a member of the 
Virgo Cluster (Bingelli, Popescu \& Tammann 1993).   It is 1.8\deg~or 1.2 Mpc in projection 
from the center of the cluster.  
Based on figure 1 of Solanes et al. (2002),
IC 3061 is in a region of relatively low HI deficiency compared to other parts of the Virgo cluster.  This
makes sense considering IC 3061's regular HI structure.

\subsection{UGC 7697 and the Ton 1542 Sightline}

Figures 5 and 6 show the HI distribution of UGC 7697.  As in Figures 1 and 2 of IC 3061, Figure 5
is the HI data for UGC 7697 over a $25^{\prime} \times 25^{\prime}$ 
Digital Sky Survey image centered on the galaxy, and Figure 6 is the velocity map showing the  
entire field of our Arecibo observations centered on the background source Ton 1542. 
UGC 7697 has the same situation as IC 3061 in that it is at the spatial edge of the cube.  
The closest galaxies to UGC 7697 detected in the Arecibo data cube are two at approximately 1300 \kms~ in the
extreme north of the cube.  These galaxies do not have previously published redshifts and will be
discussed in Rosenberg et al. (2005).  

The properties of UGC 7697 are shown in Table 1.
The HI mass of this galaxy is 2.2 $\times 10^{9}$ \Msun~from our Arecibo data,
using a distance of 28.7 Mpc from Tully-Fisher measurements (Solanes et al. 2002).  
Haynes et al. (1999) found a total HI mass of
$1.7 \times 10^{9}$ \Msun~at this distance.  Using the Arecibo data and the same method outlined for IC 3061, the dynamical mass of 
this galaxy is $1.1 \times 10^{11}$ \Msun, where r$_{\rm HI}=35.4~h^{-1}_{70}$ kpc, $\Delta$v$=234$\kms, and $i=$ 83\deg~ (Huchtmeier \& Richter 1989).  
The sightline of the absorber is 12\arcmin~ from the optical center of UGC 7697
and 7\arcmin~ from the edge of the observed rotating HI disk.  In projection, at a distance of 28.7 Mpc, this is
$100~h^{-1}_{70}$ kpc from the optical center and $58~h^{-1}_{70}$ kpc from the HI edge 
measured at 1.2 $\times 10^{19}$ cm$^{-2}$. 
As previously stated, we are limited by our beamsize when estimating the size of the HI disk because
the Arecibo beam is $30~h^{-1}_{70}$ kpc at 28.7 Mpc.

Figures 6-8 show that the velocity of UGC 7697 extends from 
$\sim$2410 \kms~to $\sim$2660 \kms.  The absorber with the closest velocity to this galaxy 
is at 2563 \kms~and, given its position on the side of the galaxy with 2410 \kms, is 
not following the gradient of rotation of the galaxy if you extend the disk beyond the HI emission
detectable here.  The properties of the absorption line are shown in Table 2.   
The column density of the single component absorber is $10^{14.3}$ cm$^{-2}$, while the column density limit
of the HI observations is $10^{18}$ cm$^{-2}$ (5$\sigma$ for a 5 \kms~channel).  The closest absorber
in velocity to the absorber at 2563 \kms~is at 1895 \kms, a difference of almost 700 \kms (Penton et al. 2004).
The  projected position of the absorber is 9\deg~from an extension of the major axis of the galaxy as shown in
Figures 5 and 6.
 Figure 7 is a right ascension-velocity map through UGC 7697 at the declination of the absorber and also shows
that the velocity of the absorber is comparable to velocities on the far side of the disk rather than on the near side. 
Table 3 presents the spatial and kinematic differences between UGC 7697 and the nearby absorber.

Figure 8 shows the integrated HI spectrum of UGC 7697 is another 
double-horned profile indicative of its edge-on nature and galaxy type (Scd).
UGC 7697 is not considered part of the Virgo Cluster, although it is not far from the extreme northern part of the cluster (7.9\deg~or
3.9 Mpc projected distance from the center).
This galaxy and absorber sightline were also examined by 
C05 and VLA maps were presented which confirm the results shown here of a smoothly
rotating disk.  
The mass we obtain from the Arecibo data is 
1.84 times the mass they obtained from the VLA, most likely representing the extended flux missed in the VLA observations.

\section{Discussion}

In both of the cases presented here the gas detected in absorption within $125~h^{-1}_{70}$ kpc of the center of the galaxy 
is not rotating with the galaxy.  
This result indicates that although dark matter halos are
presumed to extend out to $\sim 200$ kpc in these types of spiral galaxies (e.g., using satellites and halo stars, 
Kochanek 1996; Prada et al. 2003: or lensing results, Koopmans, de Bruyn, \& Jackson 1998), the gaseous baryons
and dark matter are not directly linked at large radii.  Also, though we commonly use the gas to trace the dark 
matter content of galaxies (e.g., de Blok \& Bosma 2002; Simon et al. 2005), at some radius this no longer holds.  
What is that radius and what dictates when the gas and dark matter become tightly coupled?

Figure 9 represents three possible scenarios for the relation of the gas detected in absorption to the galaxies.
The gas could be in the halo and counter-rotating, in the halo and infalling, or in
the halo and outflowing.
The final possibility is that the gas
detected in absorption is not related to the galaxies' halos and simply traces a background/foreground 
cosmic filament.  These possible scenarios for the relationship 
between the galaxies and absorbers are outlined below and the questions with regard to a dark matter$-$baryon link at large
radii are also discussed.

\subsection{Galaxy Fuel?}

The idea of the IGM feeding galaxies with star formation fuel has been developed in several models since
the initial paper on the relationship between cooling gas and galaxy formation by White \& Rees (1978) 
(e.g.  Murali et al. 2002; Yoshida et al. 2002; Maller \& Bullock 2004 and references within
these papers).  The models typically have cool clouds at distances up to $\sim 150$ kpc from
the center of the galaxy, beyond the projected radii of the gas traced by the absorbers discussed here. 
The shock heated gas is decoupled
from the dark matter as it comes into the galaxy, but must reach equilibrium with the rotating halo
during the accretion process.  It is possible the gas detected here has been recently accreted and is still
coming into equilibrium with the rotating dark matter halo, but how does the gas begin to condense
if it is counter-rotating with respect to the galaxy?

In the case of our own Galaxy, the high-velocity clouds (HVCs) represent clouds of gas that do not
follow a simple model of Galactic rotation and have been proposed to represent condensing,
infalling gas (e.g., Maller \& Bullock 2004).   The HVCs have higher column densities 
(typically $10^{18-20}$ cm$^{-2}$) than the
gas traced here, but a comparison of the relative kinematics is useful if one takes into account
the velocity component being measured for the absorbers is parallel to the galaxies, while the velocity 
component measured for HVCs is towards the Galaxy at the position of the Sun.   The two systems studied here have the absorber roughly
along the major axis of the galaxy and the infall scenario of Fig. 9 can be examined to understand the potential relationship
between the absorber and the galaxy.  For IC 3061, assuming the rotation curve remains flat out to $122~h^{-1}_{70}$ kpc, the
gas traced by the absorber deviates from galactic rotation by -360 \kms.  In the case of UGC 7697 the gas
traced by the absorber deviates by +153 \kms~from an
extension of flat rotation out to $100~h^{-1}_{70}$ kpc.  However, one should also consider the absorbers are not exactly along the plane 
of the galaxies, and this may affect the expected rotation.
Recent work on the rotation of gas at high latitudes in spiral galaxies has shown that as you move into the halo, the gas
is rotating more slowly, typically on the order of 25-50 \kms~slower at heights several kpc above the plane (e.g., Swaters et al. 1997; 
Barbieri et al. 2005; Chaves \& Irwin 2001).  Taking this decrease into account, the maximum effect possible
is for the gas to stop rotating such that it has the systemic velocity of the galaxy at higher latitudes.   Therefore the minimum
deviation between the velocities of the absorbers and the expected rotation of the gas is 32 \kms~for UGC 7697 and 202 \kms~for
IC 3061 (see Table 3).  

The velocity component measured for the absorption line 
system (V$_{\rm los}$), the unknown velocity
component tangential to our line of sight (V$_{\rm tan}$), and the potential infall
velocity into the galaxy (V$_{\rm infall}$) for the case of IC 3061 are shown in Figure 9.  For UGC 7697,
the galaxy is rotating in the opposite direction and the absorber velocity measured (V$_{\rm los}$)
is still counter to the galaxy's rotation, so for the infall scenario the absorber would be in
front of the galaxy, rather than behind.  Figures 10 and 11 show the magnitude of the tangential velocities and infall
velocities implied by this infall scenario for the absorbers near IC 3061 and UGC 7697 respectively.
If the gas traced by the absorbers is
behind IC 3061 or in front of UGC 7697, the tangential velocity component could
potentially change the counter-rotating nature of the absorbers into infalling gas clouds.  
This infall velocity can also be compared more directly to the velocity measured
for HVCs towards part of our Galaxy.

Figure 10 shows the relationship between
the tangential velocity (V$_{\rm tan}$), the infall velocity (V$_{\rm infall}$), and the distance (from
the galaxy) to the absorber for IC 3061 for the absorbing gas to no longer be counter-rotating.  
This plot uses the magnitude of the absorber velocity along the line of sight relative to the galaxy.
Since as discussed above, the gas may be rotating more slowly at higher latitude than in the main disk of the galaxy,
we use the minimum possible difference between the expected velocity of gas at that radii and
the velocity of the absorber, or the difference between the velocity of the absorber and the
systemic velocity of the galaxy.
Fig. 10 shows that with the magnitude of the absorber's velocity component along the line of 
sight relative to the galaxy at $|$V$_{\rm los}$$|$ = 202 \kms, the infall
velocity would be greater than 250 \kms~for all distances less than $200~h^{-1}_{70}$ kpc from the galaxy center.  
This is comparable to some of the very high velocity HVCs relative to the Galaxy (the maximum
is approximately $|$V$_{\rm GSR}$$|$ = 300 \kms; Putman et al. 2002), and suggests infall is
a possibility for the gas traced by the absorber.  However, we note that many HVCs
are expected to be closer than 150 kpc (e.g., Maloney \& Putman 2003; Putman et al. 2003b;  Wakker 2001; Maller
\& Bullock 2004), 
and placing the gas traced by the absorber at distances below $150~h^{-1}_{70}$ kpc from IC 3061 requires very 
high infall velocities ($> 350$ \kms).  

Figure 11 shows the same relationship between tangential velocity (V$_{\rm tan}$), infall velocity (V$_{\rm infall}$), 
and distance to the gas cloud for UGC 7697.  This gas cloud could more easily by infalling than the
case of IC 3061.  The difference between the velocity of the absorber and the systemic velocity
of the galaxy is only 32 \kms, so with this $|$V$_{\rm los}$$|$ if the cloud is at $150~h^{-1}_{70}$ kpc, the tangential 
velocity of the cloud would only have to be greater than 29\kms~to have infall and no counter-rotation.  
The infall component to the galaxy would then be 43\kms.  In fact, all of the possible distances
from the absorber to UGC 7697 are consistent with the possibility of infall and comparable to the 
velocities of HVCs relative to the Galaxy.  However, this gas is also much further away (at least 100 kpc) than the
distances to many HVC complexes (e.g., Wakker 2001; Putman et al. 2003b).  Both of these absorbers
may potentially represent the cooling, infalling halo gas before it has condensed to the densities of HVCs.

\subsection{Galaxy Waste?}

The case of the gas traced by the absorbers being galactic waste from IC 3061 and UGC 7697 is difficult to 
reconcile given their positions roughly along the plane of both galaxies.  Galactic worms, mushrooms, and
chimneys in spiral galaxies represent the possible 
blow out of enriched gas into the halo of a galaxy (e.g., English et al. 2000; McClure-Griffiths et al. 2003), but
are almost always seen extending up in galactic latitude.  These features may be due to supernovae at high galactic latitude, where
the nature of the diffuse material at high latitude allows the explosion to push through the main disk of the galaxy.  In addition
models predict outflowing gas due to multiple supernovae would subsequently fall back onto the galaxy closer to 
the galactic center (e.g., Bregman 1980), and
the outflowing gas itself is not likely to reach distances greater than 10 kpc from the galaxy's disk (de Avillez 2000).
There is no known scenario where blown out material would start counter-rotating in the plane of the galaxy.
For both of the galaxies studied here, the absorbing gas is at distances greater than $50~h^{-1}_{70}$ kpc from the edge of the galaxy's
HI disk, an unlikely location to find outflowing material from a spiral galaxy.

The last panel of Figure 9 shows the outflow scenario for IC 3061.
For UGC 7697 the gas would be beyond the galaxy in the outflow scenario.
In Figures 10 and 11 we can turn infall velocity to outflow velocity to examine the magnitude of the velocity 
component needed to have the gas outflowing and not counter-rotating.  For IC 3061 the outflow
velocity (within $200~h^{-1}_{70}$ kpc) would have to be greater than at least 250 \kms~for the gas to be outflowing 
and not counter-rotating, and for UGC 7697, the outflow velocity would have to be greater than at least 37 \kms.
Given the absorbers' distances and positions along the major axes of the galaxies, away from active
star formation in the galaxy disks and off axis from any outflow from the galaxy centers, these
outflow velocities seem unlikely and the galactic waste model does not seem to be a plausible explanation
for the origin of the gas.  The only remaining galactic waste scenario is a low surface brightness satellite galaxy which has
blown out most of its gas due to multiple supernova (Stocke et al. 2004) or the presence of a structure representing
the tidal destruction of a satellite, similar to the Magellanic Stream around our own Galaxy (e.g., Putman et al. 2003), but more diffuse, as the Magellanic Stream would have
been detected by our deep HI observations.   The next section discusses this scenario further.

\subsection{Galaxies with Counter-Rotating Halo Gas and Galactic Satellites}

It is possible these two systems are anomalies, like the few examples of galaxies with counter-rotating
gas in emission (e.g., Hunter et al. 1998;  Pizzella et al. 2004).
This is normally attributed to the recent accretion of a satellite, or merger.  There is no evidence for a
recent interaction in the galaxies' HI distributions to the deep levels of these observations (Figs. 1-8).
A typical cloud in the Magellanic
Stream has a linewidth of $\sim 25$ \kms~ and a column density above $10^{19}$ cm$^{-2}$ 
(Putman et al. 2003a).  
Our 5$\sigma$ column density sensitivity to this type of cloud is 
$6 \times 10^{18}$ cm$^{-2}$ (IC 3061) and $5 \times 10^{18}$ cm$^{-2}$ (UGC 7697).  Our beamsize at the distance of IC 3061 is $39.6~h^{-1}_{70}$ kpc 
and $29.8~h^{-1}_{70}$ kpc for UGC 7697, while the Magellanic Stream subtends over 100\deg~on the sky, 
or $\sim$100 kpc if you place the entire feature at 55 kpc from the Galaxy.  Therefore a feature
such as the Magellanic Stream or a feature $1/3$ the size and twice as diffuse would be detected by
these observations.
There is also no evidence for a recent interaction in the optical DSS images of the galaxies (Figs. 1 \& 4).  The closest
galaxy to IC 3061 with a velocity within 750 \kms~of IC 3061 is $390~h^{-1}_{70}$ kpc away.  
For UGC 7697, the closest galaxy within 750 \kms~is $344~h^{-1}_{70}$ kpc away.
A counter-rotating low mass
satellite in the plane of the galaxy is the only remaining possibility, and an unlikely one for several
reasons, including the following.  1.  The mass limit of our HI observations is 7.5 $\times$ 10$^{6}$ \Msun~for IC 3061
and 3.6 $\times 10^{6}$ \Msun~for UGC 7697 (5$\sigma$; $\Delta$v = 50\kms) and thus any
gas rich satellite galaxy would have been detected.  For UGC 7697 this is further supported by the higher
resolution VLA observations of C05.  
2.  Satellites are very rarely found orbiting a galaxy in the plane
of the galaxy, and are most commonly found on polar orbits (e.g. Holmberg 1969; Zaritsky et al. 1993; Knebe et al. 2004). 
 
\subsection{A Gaseous Local Filament}

Both of the absorbers discussed here
have velocities similar to a nearby galaxy (within 210 \kms~of the systemic velocity), but
their relationship to the actual kinematics of the galaxy is somewhat dubious.
A remaining explanation for the gas traced by the absorbers is that it does not have a 
direct connection to the gas in the galaxy.  Instead, the association between the galaxy and the absorber 
could be a general connection with the large scale structure of the region.
Simulations show that a substantial fraction of the low column density gas in the vicinity of galaxies connects 
with large scale structure, and not necessarily with individual galaxy halos (Dav\'e et al. 2001).

Observationally, Bowen et al. (2002) found that within 37-290 $h^{-1}_{70} $ kpc of a galaxy 
the covering factor is $\sim 100$\% for low column density gas (N$_{HI} > 10^{13}$ cm$^{-2}$). 
Nevertheless, they were unable to definitively associate the gas seen in absorption directly with a single galaxy halo 
because of the complexity of the galaxy distribution in the region. A connection between absorbers and 
large-scale structure was concluded by Penton et al. (2002) by performing a 
correlation function analysis on a sample of galaxies and absorbers.  They found that absorbers with column 
densities in the range of those studied here are more weakly clustered with galaxies than galaxies are clustered 
with each other. They also found statistical evidence that absorbers are associated with large scale filaments, as also
found by Rosenberg et al. (2003).
The kinematic and spatial relationships between the galaxies and absorbers studied here and in the sample of C05, provide
additional evidence for low column density absorbers probing gas associated with large scale structures rather than
individual galaxy components.
The gaseous material traced by the absorbers may ultimately fall into the galaxy halos and fuel future star formation, but currently 
the gas does not follow the galaxies' potential.

Both absorption line-galaxy pairs in this sample reside in gas-rich and galaxy-rich regions of space,
providing further support for the idea that the absorbers are tracing out a cosmic filament.  IC 3061 is on the 
outskirts of  the Virgo Cluster, where gas-rich galaxies and absorbers are very common (Solanes et al. 2001; 
Rosenberg et al. 2003). While UGC 7697 does not fall within the bounds of the Virgo Cluster, it is not far from 
the boundary and is probably still in a significantly over-dense region.  Figure 12 shows the sightlines towards
PG 1211+143 and Ton 1542 and galaxies from the Updated Zwicky Catalog (Falco et al. 1999)
over a 3500 \kms~velocity range including the absorbers and galaxies of interest here.
The absorber near IC 3061 along the sightline to PG 1211+142 is clearly in a denser region than 
the absorber near UGC 7697 along the sightline to Ton 1542, but clearly both are near a filament of galaxies.  
In the sample of absorbers considered here 
and in C05, the UGC 7697 - Ton 1252 pair is actually the most isolated when a $\sim$2 Mpc radius around the absorber is
examined.   In another words, all of the galaxy-absorber pairs lie along cosmic filaments rich in galaxies and 
intergalactic gas. 
It may be that the HI gas in galaxies is a better tracer of the location of gaseous cosmic filaments than the stellar 
component of galaxies.  This would be true if much
of the HI gas represents recently accreted star formation fuel from the cosmic filament.  The link between a large
sample of absorbers and HI galaxies will be addressed in
future papers (Rosenberg et al. 2005, Ryan-Weber 2005).

\subsection{Link between Gas and Dark Matter}

At what radii would low column density gas traced in absorption directly trace the potential
of the nearby galaxy and dark matter distribution?   We commonly attempt to detect the 
gas of a spiral galaxy out to larger and larger radii with deep HI or H$\alpha$ observations to probe their dark matter
content (e.g., Corbelli, Schneider, \& Salpeter 1989; Bland-Hawthorn, Freeman \& Quinn 1997), but our results indicate
there is a maximum radius at which this method is effective.  The idea of a link between the surface
density of gas and dark matter was discussed by Hoekstra, van Albada, \& Sancisi (2001; see also Bosma 1981). 
They derived a mass model and applied it successfully to a sample of 24 spiral galaxies, but were unable to find
real evidence for a coupling between HI and dark matter in spiral galaxies.  Our results 
indicate that if the gas and dark matter are coupled, or even if the gas simply follows the mass distribution of the galaxy,
there is a maximum distance or minimum density for this relationship.  The absorbers studied here probe beyond this limit.

Both IC 3061 and UGC 7697 are spiral galaxies similar to the Milky Way in luminosity and HI mass and thus
a reasonable total mass (consisting primarily of cold dark matter) is $\sim10^{12}$ \Msun~(e.g., Battaglia et al. 2005; Moore et al. 2001).
By examining the cold dark matter models of Navarro, Frenk, \& White (1996), one can estimate the
dark matter density at the radii of the absorbers.  At 100 or 122 kpc from the center of a $\sim10^{12}$ \Msun~
dark matter halo, the dark matter density would be $\sim 60,000$ \Msun~kpc$^{-3}$ and the material
would be on the flat part of the rotation curve.  Clearly the dark matter would dominate over the gas in mass at this
radii, even for an absorber that is 100 kpc in diameter.
One option to explain the lack of rotation in the gas detected here
is a truncation of the dark matter halo due to tidal stripping (Nagai \& Kravtsov 2005).  This tidal stripping
would result in a lack of both dark matter and gas rotating at large radii.  
In the cluster simulations of Nagai \&
Kravstov, they find the average mass loss of sub-halos is $\sim$30\% near the virial radius of the cluster.  IC 3061 is considered
part of the Virgo Cluster and in projection is 1.2 Mpc from the center of the cluster.
UGC 7697 is not considered part of the cluster and is 3.9 Mpc in projection from its center.  Both galaxies are $\sim1000$ \kms~
from the velocity centroid of the Virgo Cluster, but could be considered as being on the upper edge of the velocity range
of the cluster (e.g., Bingelli et al. 1987).
The virial radius of the Virgo Cluster has been estimated
to be 1.61 $h^{-2/3}_{70}$ Mpc (Mamon et al. 2004), so IC 3061 is a candidate for the tidal truncation scenario, but
UGC 7697 does not fall naturally into this model.    In support of the truncation model is the recent work on the dark matter halo of the Milky
Way by Battaglia et al. (2005), which indicates a truncated model may be necessary to explain the velocity distribution
of objects out to 120 kpc from the Milky Way.  Lack of support for this scenario come from the recent results of Prada et al. (2003), who 
examined 3000 satellite velocities from the SDSS and found that galaxies are embedded in large halos extending out to 350 kpc.
The regular structure of both galaxies' HI and stellar disks also defies the scenario of a 
severe tidal disruption.  Finally, another possible model is that the dark matter and baryonic halos of the galaxies have slightly different shapes, i.e.,
the dark matter halo is oblate, while the baryonic halo is spherical or otherwise (e.g., Bailin et al. 2005).
 
The result of low column density gas not smoothly rotating with a dark matter halo may represent
a limit on the density of gas that is able to come into equilibrium with a rotating galaxy.
This could be due to other forces, besides gravity, that act on diffuse gas more
readily than gas above $\sim10^{19}$ cm$^{-2}$ in HI column density.
Ram pressure forces are the obvious example (e.g., Kenney, van Gorkom \& Vollmer 2004), and since these galaxies are moving, this diffuse
component may not be able to hang on.  Ram pressure stripping for the origin of the gas for these
galaxies seems unlikely however, as the counter-rotating nature of this gas indicates it
was not recently stripped from the galaxy and both galaxies have a regular HI structure.\footnote{In a sample 
of Virgo galaxies imaged in HI by the VLA, the closest galaxy to IC 3061, NGC 4189, also does not show signs of ram pressure
stripping (Chung et al. 2005).}
In the case of the Milky Way, there may be a similar effect of low column density gas not able to follow the
dark matter halo as closely.  The HVCs have typical peak column densities of just below
$10^{19}$ cm$^{-2}$ (Putman et al. 2002), and most of the HVCs with peak columns between $5 \times 10^{19}$ and
$10^{20}$ cm$^{-2}$ (the maximum column density observed for an HVC in the southern sky) can be attributed
to the Magellanic Stream, an HVC which is known to originate from the interaction
of the Magellanic Clouds with the Milky Way.  If some of the lower column density HVCs represent the condensing IGM, the reason they do not
appear to be rotating with the Galaxy may be related to their relatively low column densities.
Perhaps diffuse IGM gas must reach a certain column density before it is able to come into equilibrium with
the dark matter halo of a galaxy.  This comparison cannot be taken too far, as the gas we are probing is at much
lower column densities than HVCs and the velocity components being measured are different (see Gaseous Infall section).

The lack of kinematic correlation between the gas and dark matter at large radii can
be combined with the work of C05 to strengthen the scenarios discussed here.  The
strongest case of this lack of correlation lies in the two systems presented here with the edge on nature
of the galaxies and absorber lying close to the plane of the galaxy at a distance of $\sim100~h^{-1}_{70}$ kpc.  In the
C05 paper, 5 additional spiral galaxies with kinematic information were found to have low column density (10$^{13.2-13.9}$ cm$^{-2}$) \lya
absorbers within $182~h^{-1}_{70}$ kpc.  These galaxies have various inclinations and absorber positions.
C05 fit these galaxies with an extended rotating gaseous disk to determine if the absorption line velocity agreed with that
expected from a Navarro, Frenk \& White (1996) cold dark matter
halo model.  For 3 of these, they could explain the rotation if one invoked warps or thick rotating gas layers, and in
the other 2 
they found the magnitude of the velocity was too high to be fit by any reasonable galactic features, or the gas would 
have to be counter-rotating.   
Since determining the main velocity
components of the absorbers in relation to the galaxies is not straight-forward for these systems, due to the galaxies' inclinations
and the locations of the absorbers, comments cannot easily be made on the issues of
gaseous inflow and/or outflow.  In summary, the C05 results support the results we present here that the gas and dark matter
do not appear to be linked in the extended dark matter halos of spiral galaxies.

\section{Summary}

The deep HI observations around low redshift \lya absorbers presented here have revealed two nearby, 
edge-on spiral galaxies within $125~h^{-1}_{70}$ kpc (in projection from the galaxies' centers) of absorption line
systems at similar velocities. 
Both of the absorbers are located roughly along the plane of the nearby galaxy and, based on the kinematics
of the galaxies revealed by the HI observations, are not rotating with the galaxy.
In the case of IC 3061 the gas traced by the absorber (along the sightline to PG 1211+143) is either
counter-rotating or infalling at a very high relative velocity.  For UGC 7697 the gas traced by the absorber (along the sightline to Ton 1542)
may be infalling to the galaxy, but cannot be rotating with the galaxy, as also found by C05.
These results indicate that despite dark matter halos extending out to 100's of kpc, there is not
an associated diffuse baryonic component bound to this dark matter at large radii.  
The results agree with previous findings that indicate low column density absorbers, even in
the vicinity of galaxies, largely trace the cosmic web.

In terms of the questions posed in the Introduction, the following "answers" hold based on this paper.
\begin{itemize}
\item Do the \lya
absorbers near galaxies trace gas in extended dark matter halos?  No, unless for some reason the gas has not
yet reached equilibrium with the dark matter.

\item   Is there gas rotating with the dark 
matter halos of galaxies at large radii?  Clearly not unless our understanding of the kinematics of dark matter halos at large 
radii is incorrect.  Another possibility is the dark matter halo has been truncated and neither gas or dark matter
is rotating at large radii.

\item Do the \lya absorbers near galaxies
represent galactic waste?  No, the gas is unlikely to have originated from the galaxies given the discrepant velocities and positions
along the major axes of the galaxies.

\item Do the \lya absorbers near galaxies represent galactic fuel?  Possibly, but in one case infall velocities are quite
high given the distance from the galaxy and a comparison to the velocities of HVCs.  The absorbers may
represent infalling gas that will eventually condense into HVCs.

\item  Are the \lya absorbers near galaxies simply part of large scale cosmic filaments?  Yes, this is the 
most likely scenario given the geometry and velocities of the absorber-galaxy pairs discussed here.  This is 
consistent with the observations of C05, Bowen et al. (2002), and Penton et al. (2002), as well as the simulations 
of Dav\'e et al. (2001).

\item Do the \lya absorbers represent the presence of a low surface brightness galaxy, rich in gas, but not
easily detected in the optical?  No, our observations reach HI masses on the order of $3.6 - 7.5 \times 10^{6}$ \Msun~
and column densities of $5-6 \times 10^{18}$ cm$^{-2}$, so an HI-rich dwarf galaxy or stripped gaseous feature
such as the Magellanic Stream would have been detected.

\end{itemize}

\medskip
\noindent
\acknowledgements{{\bf Acknowledgements: }We would like to thank E. Ryan-Weber, M. Shull, J. van Gorkom and 
J. Bullock for useful discussions and the staff at Arecibo for all of their assistance, in particular Phil Perillat.
We also thank the referee for useful comments.  
MEP acknowledges partial support by NASA through grant HST-HF-01132.01 awarded by STScI.
JLR acknowledges NSF Grant AST-0302049.  JTS and JLR thank HST project awards AR-09221.01-A and GO-06593.01-A
for support.
This research has made use of the NASA/IPAC Extragalactic Database (NED) which is operated by the Jet Propulsion 
Laboratory, California Institute of Technology, under contract with the National Aeronautics and Space 
Administration.  The Digitized Sky Surveys were produced at the Space Telescope Science Institute under U.S. 
Government grant NAG W-2166. }

\clearpage

\begin{deluxetable}{lcc}
\tabletypesize{\scriptsize}
\tablecolumns{9}
\tablecaption{Galaxy Properties
\label{tbl1}}
\tablewidth{3in}
\tablehead{
\colhead{} & \colhead{IC 3061} & \colhead{UGC 7697}}
\startdata
Class & SBc & Scd \\
$\alpha$ (J2000) & 12h15m04.51s & 12h32m51.62s \\
$\delta$ (J2000) &  +14d01m44.4s &  +20d11m01.2s \\
V$_{\rm sys}$ (\kms)& 2332 & 2531 \\
V & 14.3 & 15.0 \\
D (Mpc) & 38.0 & 28.7 \\
r$_{\rm HI}$ (kpc) & $52~h^{-1}_{70}$ & $35~h^{-1}_{70}$ \\
M$_{dyn}$ (\Msun)& $2.7 \times 10^{11}$ & $1.1 \times 10^{11}$ \\
M$_{\rm HI(arecibo)}$ (\Msun) & $5.3 \times 10^{9}$ & $2.2 \times 10^{9}$ \\
\enddata
\end{deluxetable}
 
 \begin{deluxetable}{lccccc}
 \tabletypesize{\scriptsize}
 \tablecolumns{3}
 \tablecaption{Absorber Properties
 \label{tbl2}}
 \tablewidth{5in}
\tablehead{
\colhead{Sightline} & \colhead{$\alpha$ (J2000)} & \colhead{$\delta$ (J2000)} & \colhead{V (\kms)} & \colhead{$\Delta$v (\kms)} & \colhead{log N$_{\rm HI}$}
}
\startdata
PG 1211+143 & 12h14m17.7s & +14d03m13s & 2130 &  97 & 13.76\\
Ton 1542 & 12h32m03.6s & +20d09m29s & 2563 & 49 & 14.03\\
\enddata
\end{deluxetable}

\begin{deluxetable}{lcc}
\tabletypesize{\scriptsize}
 \tablecolumns{5}
 \tablecaption{Absorber and Galaxy Relationship
 \label{tbl2}}
 \tablewidth{4in}
\tablehead{
\colhead{} & \colhead{PG 1211+143} & \colhead{Ton 1542}
}
\startdata
Nearest Galaxy & IC 3061 & UGC 7697 \\
d$_{\rm galaxy center to absorber}$ (kpc) & $122~h^{-1}_{70}$ & $100~h^{-1}_{70}$ \\
d$_{\rm galaxy HI edge to absorber}$ (kpc) & $77~h^{-1}_{70}$ & $58~h^{-1}_{70}$ \\
$\Delta$v$_{\rm absorber - galaxy systemic}$ (\kms)& $-$202 & $+$32 \\
$\Delta$v$_{\rm absorber - galaxy closest edge}$ (\kms)& $-$360 & $+$153 \\
\enddata
\end{deluxetable}

\clearpage



\clearpage

\begin{figure}
\caption{The HI distribution of IC 3061 over its 25\arcmin~$\times$ 25\arcmin~DSS image with the 
position of the background source, PG 1211+143, used to detect the Ly$\alpha$ absorption line with a similar
velocity to the galaxy labeled.  The projected distance from the absorber to the center of IC 3061 is shown.
Contours represent column density levels of 1.4, 2.8, 5.6, 11, 17, and 23 $\times 10^{19}$ cm$^{-2}$. 
The contours are clipped on the left due to the galaxy being at the edge of our observing area.
The galaxy at the lower left was also outside of the Arecibo map (see Fig. 2) and has a velocity over 2000 \kms~
below that of IC 3061.}
\end{figure}

\clearpage



\clearpage

\begin{figure}
\includegraphics[angle=-90,scale=.9]{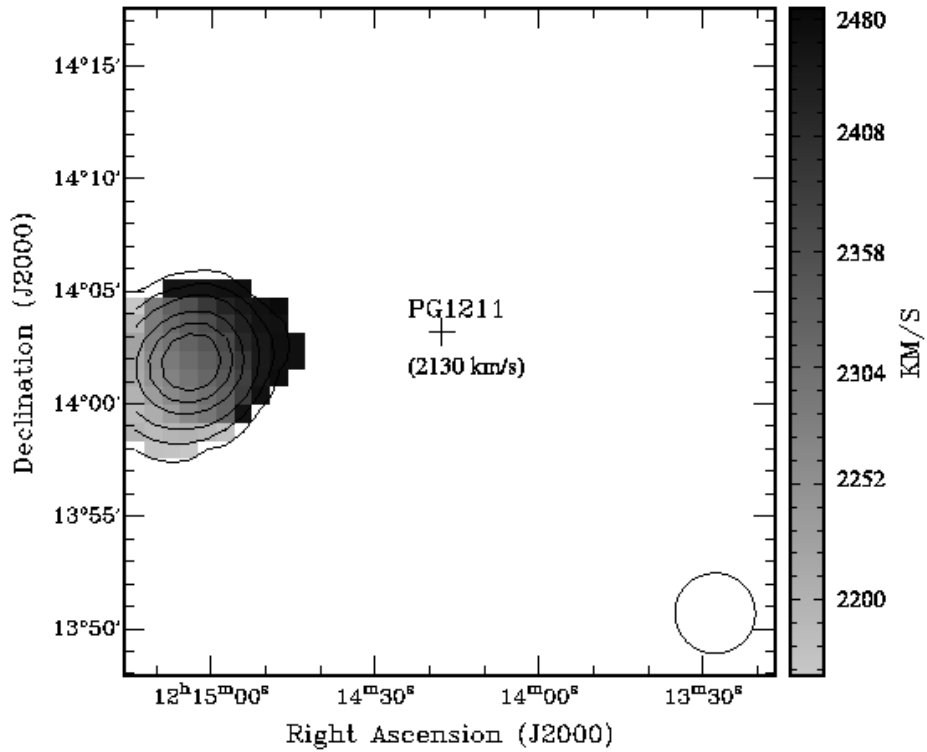}
\caption{The neutral hydrogen contours of IC 3061 (same as Fig. 1) overlaid on a greyscale velocity map. The scale bar on the right shows the correspondence between the grayscale and the velocity scale. The position of PG 1211+143 and the velocity of the absorber are labeled. The beam is shown in the lower right corner.}
\end{figure}

\clearpage


\clearpage

\begin{figure}
\caption{The neutral hydrogen contours of IC 3061 in right ascension $-$ velocity space at
approximately (given Arecibo's resolution) the declination of the absorber (noted in the upper left).   
Contours are 5, 10, 15, 20, 25, 30, 35, 40, and 45 mJy/bm. Refer to Fig. 1 or 2 for the declination slice 
of the galaxy being shown.  The position in RA-velocity space of 
the absorber towards PG 1211+143 is labeled.  Note the closest spatial position of the
galaxy to the absorber has the furthest velocity from that of the absorber.}
\end{figure}

\clearpage



\clearpage

\begin{figure}
\includegraphics[angle=-90,scale=.7]{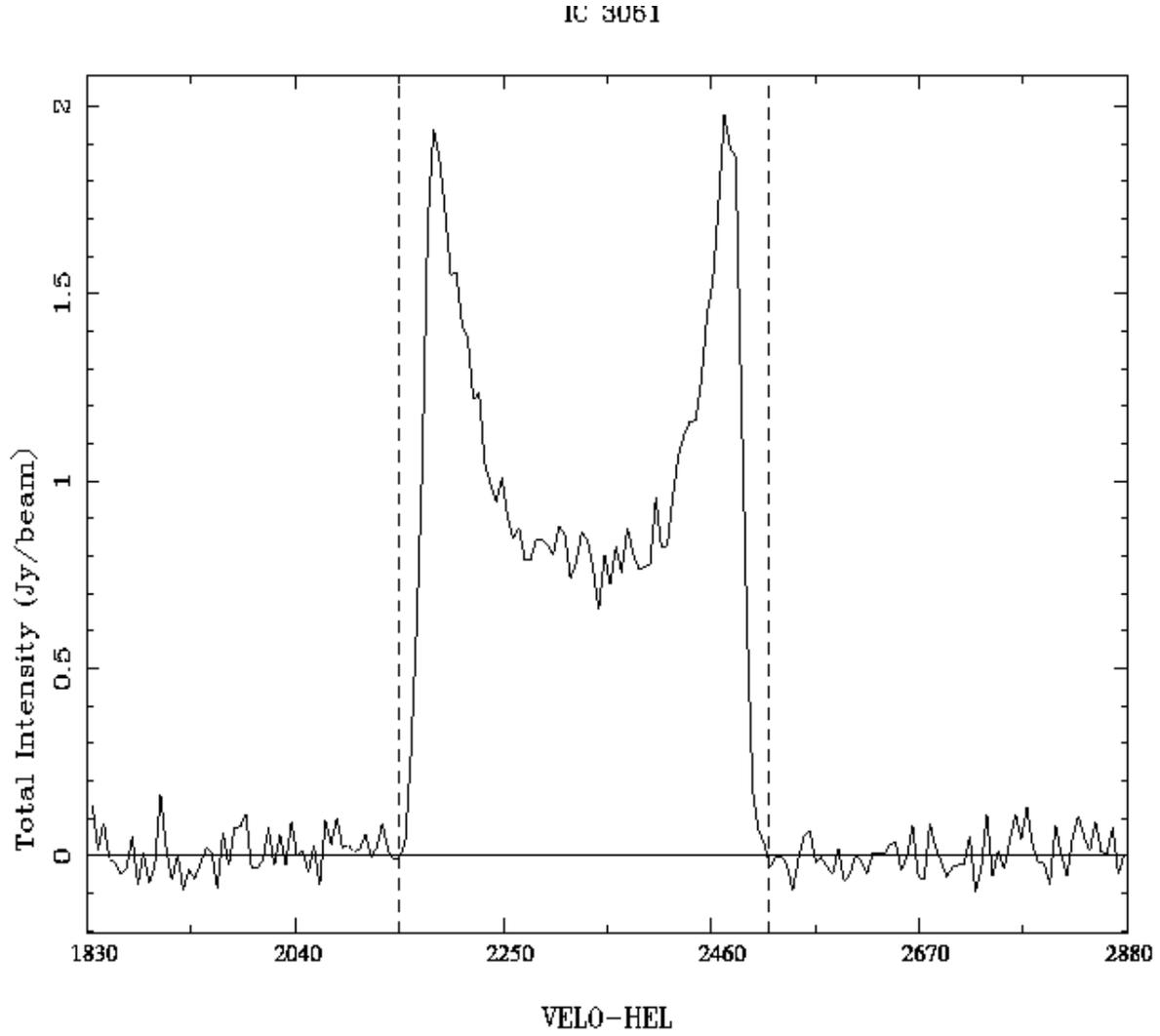}
\caption{The integrated HI spectrum of IC 3061.}
\end{figure}

\clearpage



\clearpage

\begin{figure}
\caption{The HI distribution of UGC 7697 over its 25\arcmin~$\times$ 25\arcmin~DSS image with 
the position of the background source, Ton 1542, used to detect the Ly$\alpha$ absorber at a similar velocity to
the galaxy labeled.  The projected distance of the absorber from the center of the galaxy is shown.
Contours represent column density levels of 1.2, 2.4, 4.8, 9.6, 14, and 19 $\times 10^{19}$ cm$^{-2}$.   
The cutoff of the contours on the left is due to the galaxy's position being at the edge of the observing area.}
\end{figure}

\clearpage



\clearpage

\begin{figure}
\includegraphics[angle=-90,scale=.9]{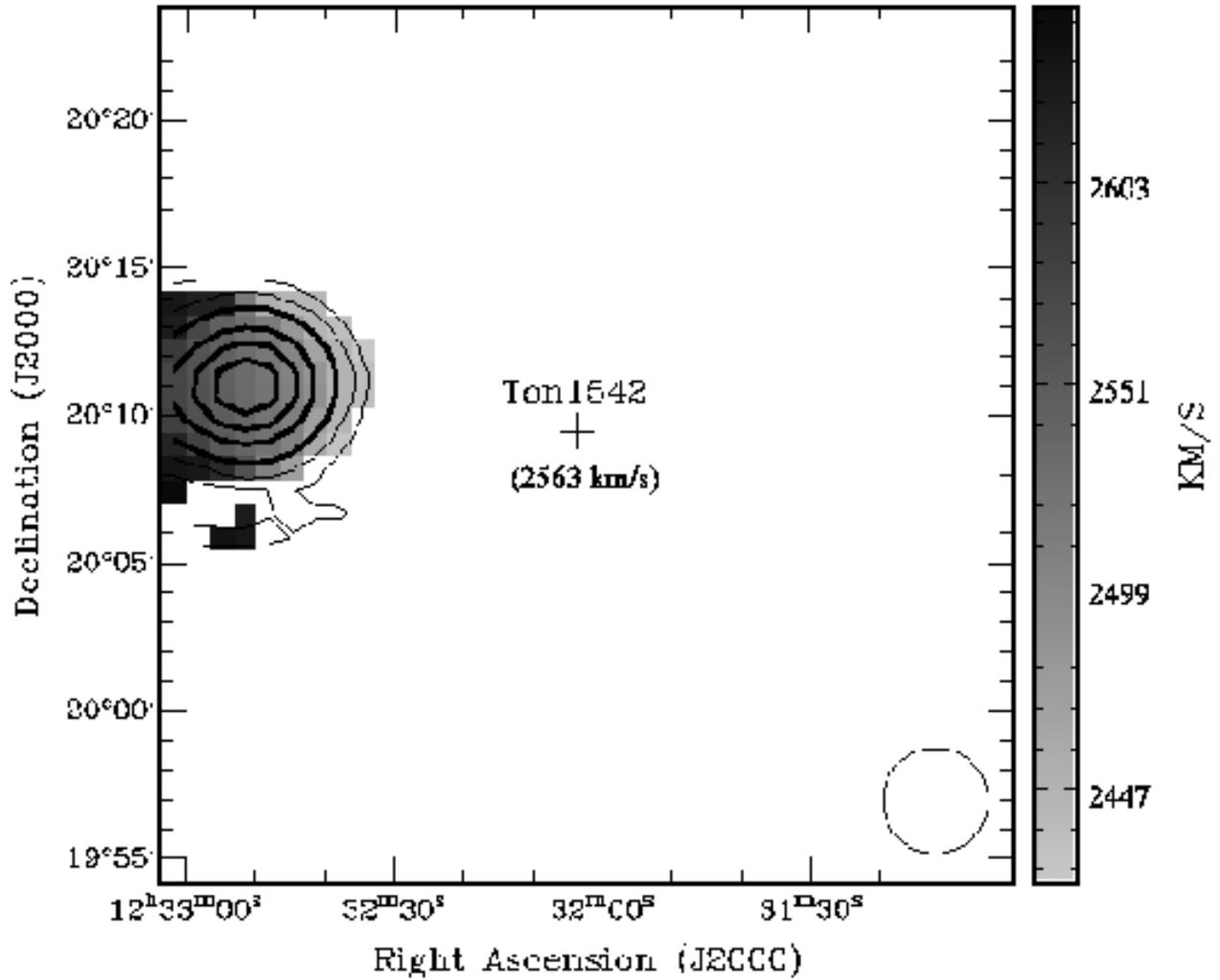}
\caption{The neutral hydrogen contours of UGC 7697 over the velocity map.  
The velocity scale is shown on the right of the image.
The position and velocity of the Ly$\alpha$ absorber towards Ton 1542 is labeled 
by the cross.  
Contours are the same as Figure 5 and the beam is shown in the lower right corner.}
\end{figure}

\clearpage



\clearpage

\begin{figure}
\caption{The neutral hydrogen contours of UGC 7697 in right ascension $-$ velocity space at
approximately the declination of the absorber (noted in the upper left).   The position in RA-velocity space of 
the absorber is labeled. 
Contours are 5, 10, 15, 20, 25, 30, 35, 40, and 45 mJy/bm.  Note the closest part of the galaxy to the
absorber has some of the furthest velocities from the absorber.}
\end{figure}

\clearpage



\clearpage

\begin{figure}
\includegraphics[angle=-90,scale=.7]{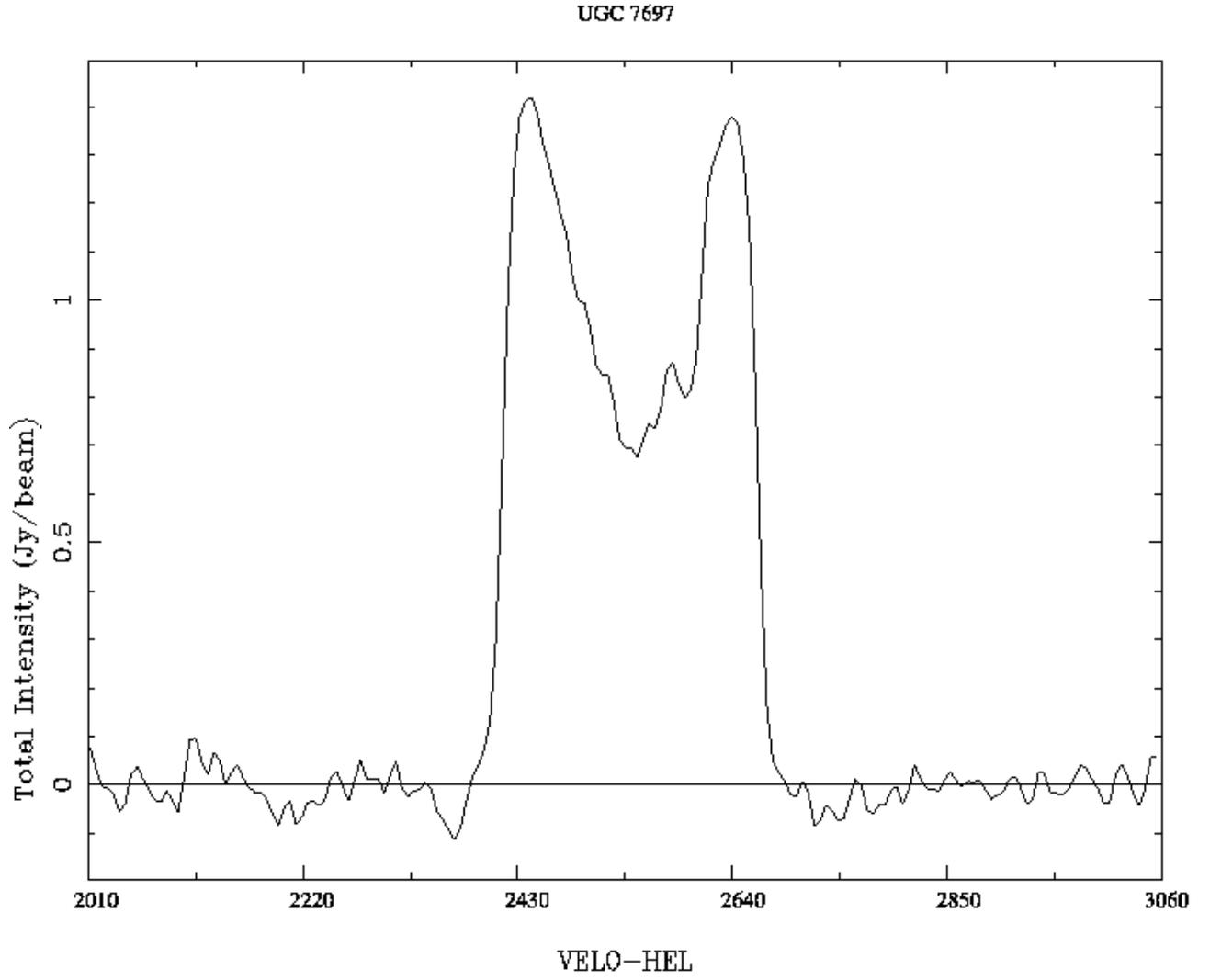}
\caption{The integrated HI spectrum of UGC 7697 with hanning smoothing.}
\end{figure}

\clearpage


\clearpage

\begin{figure}
\includegraphics[angle=0,scale=.7]{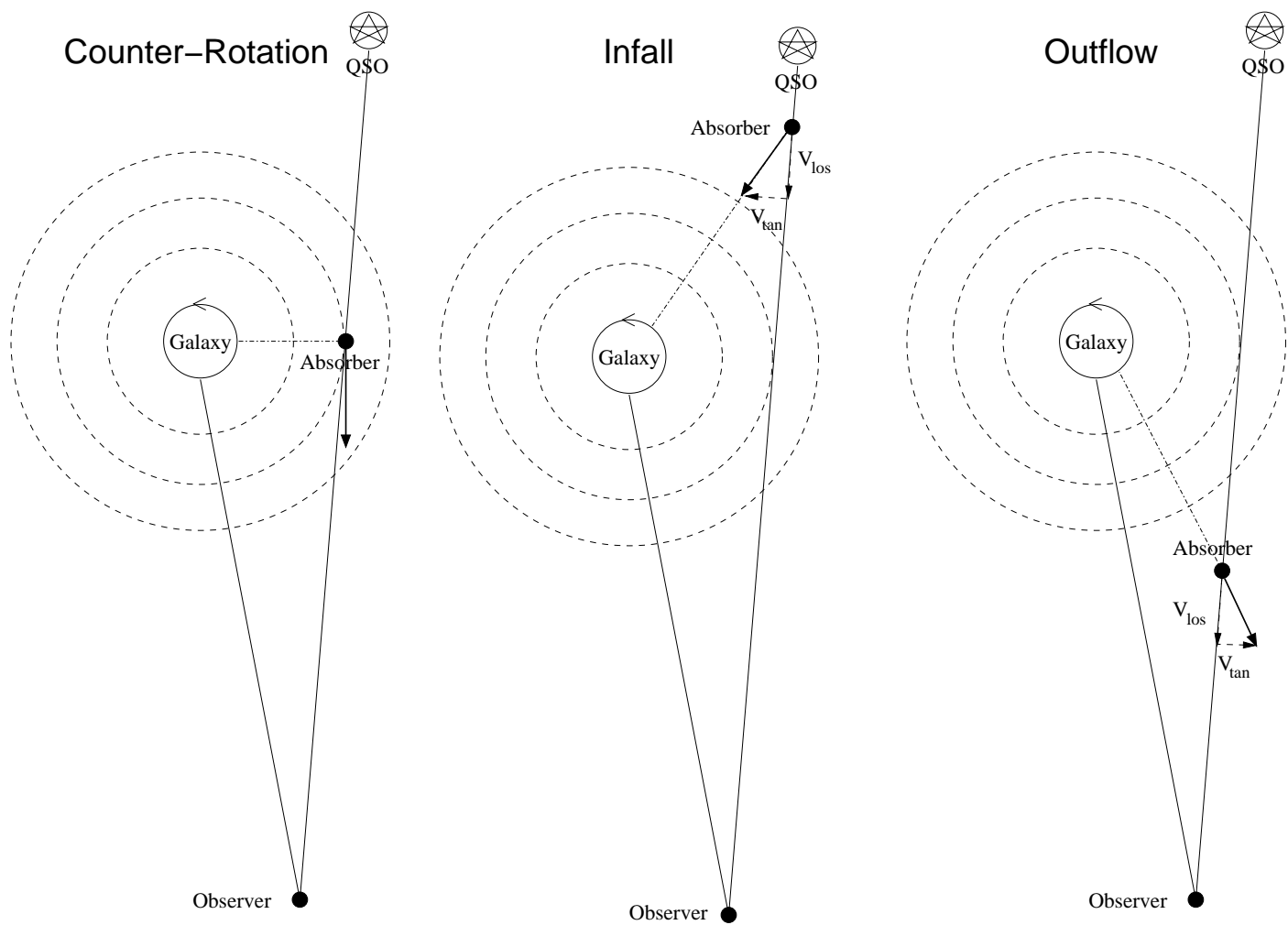}
\caption{Three possible scenarios for the relation of the gas traced by
the absorber and the galaxy in the case of IC 3061.   This figure is obviously not to scale, in particular
the background QSO used to trace the absorber would be at much greater distance from the galaxy.
For UGC 7697 the scenarios are the same but the galaxy is rotating
in the opposite direction and thus the velocity vectors of the absorption line system are also
reversed (i.e., the infall scenario would have the absorber closer to the observer, and the outflow
scenario would have the absorber beyond the galaxy).  
V$_{\rm los}$ is the velocity measured by the absorption line
observations and V$_{\rm tan}$ is the unknown tangential velocity to us.  In the middle panel
the possible infall velocity (V$_{\rm infall}$) is represented by the solid arrow towards the galaxy, where
in the outflow scenario, the potential outflow velocity is represented by the solid arrow away from
the galaxy.}
\end{figure}

\clearpage


\clearpage

\begin{figure}
\includegraphics[angle=0,scale=.6]{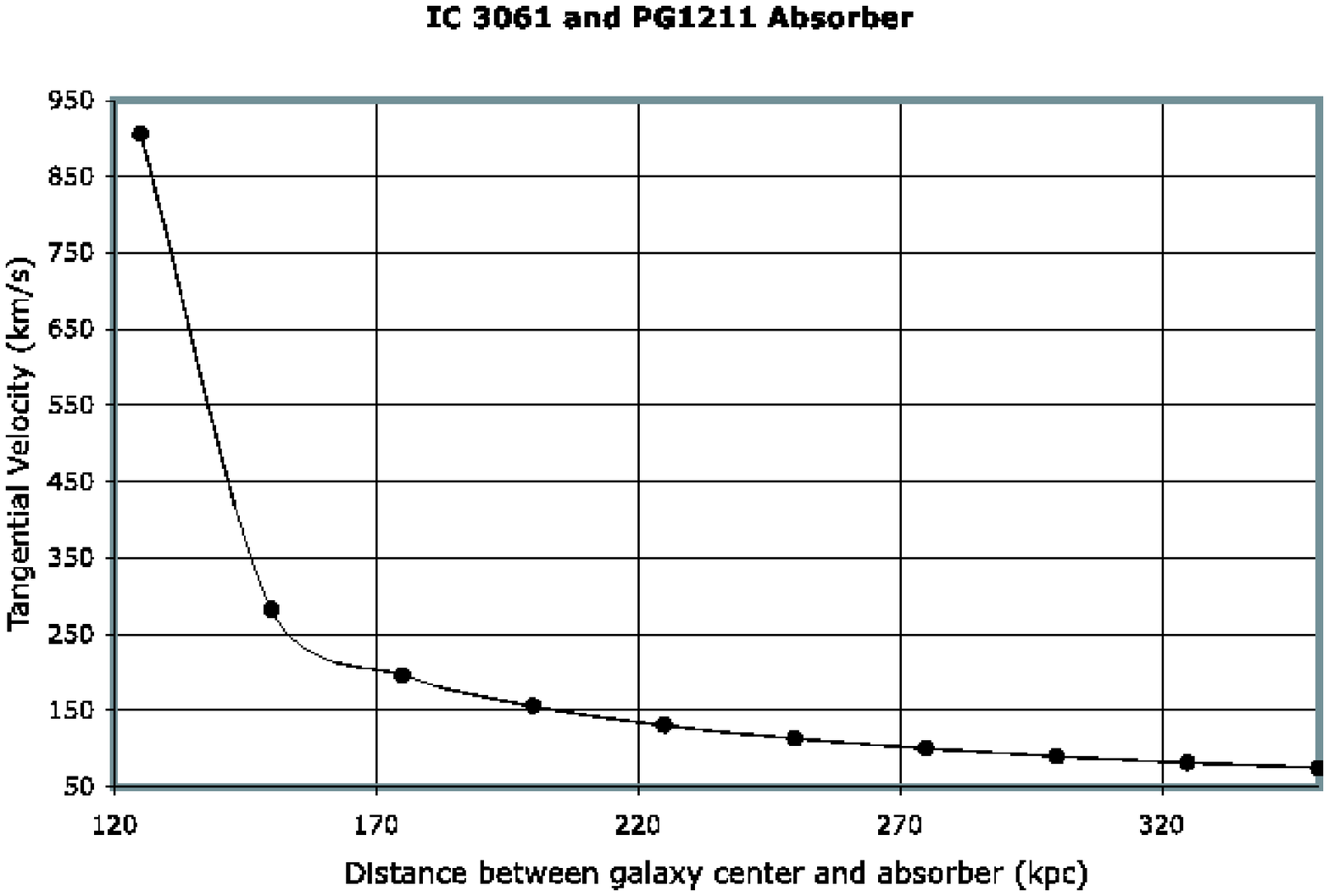}
\includegraphics[angle=0,scale=.6]{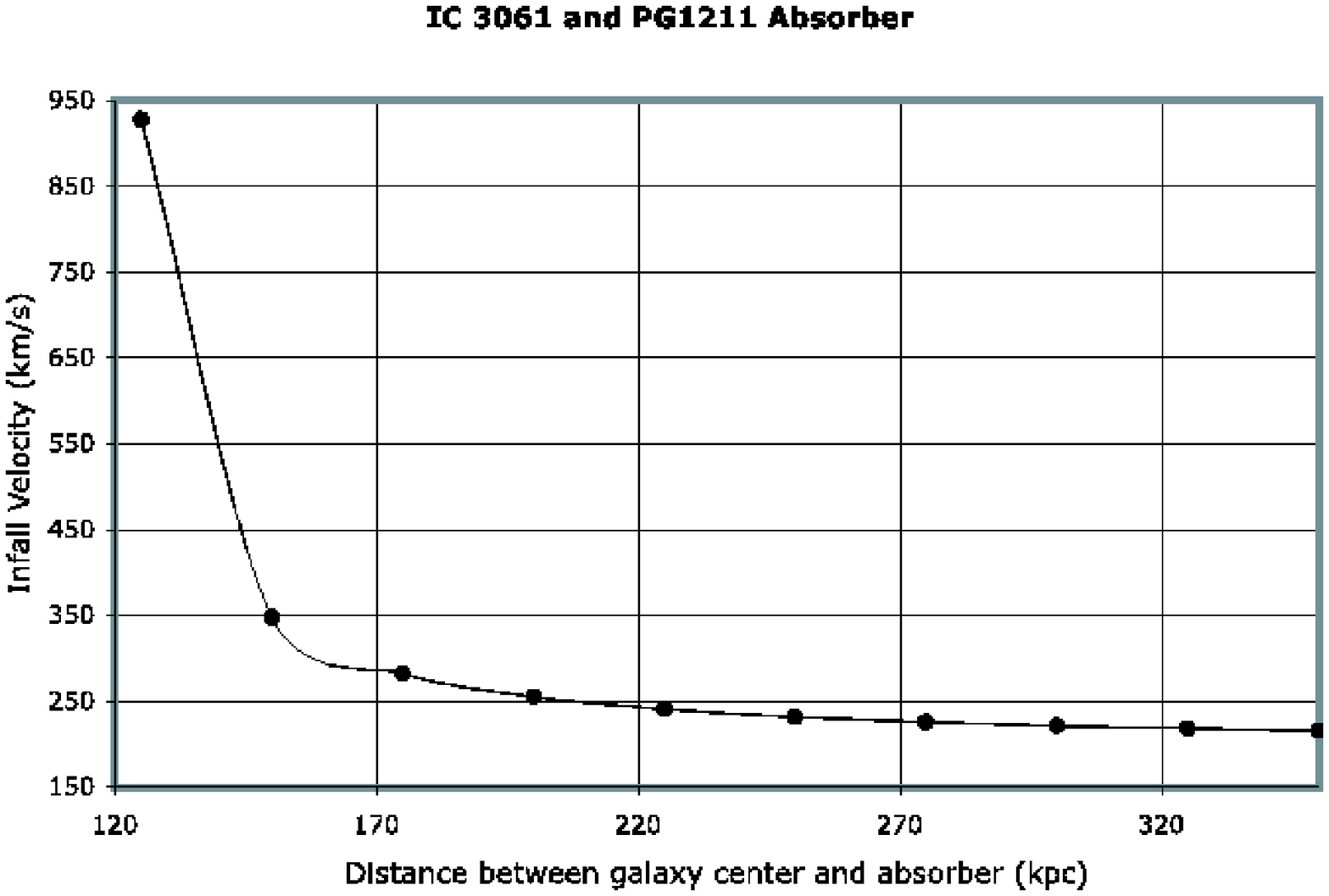}
\caption{The top panel shows the magnitude of the minimum tangential velocity at each distance from 
the center of IC 3061
for the gas traced by the absorber PG 1211 to be infalling (only if the gas is behind the galaxy) or
outflowing (only if the gas is in front of the galaxy) and not counter-rotating (see Fig. 9).  The bottom panel shows the 
corresponding magnitude of the minimum 
infall/outflow velocity.  The minimum 
magnitude of the V$_{\rm los}$ component
of the absorber relative to IC 3061 is the difference between the velocity of the absorber
and systemic velocity of the galaxy, or 202 \kms.}
\end{figure}

\clearpage


\clearpage

\begin{figure}
\includegraphics[angle=0,scale=.6]{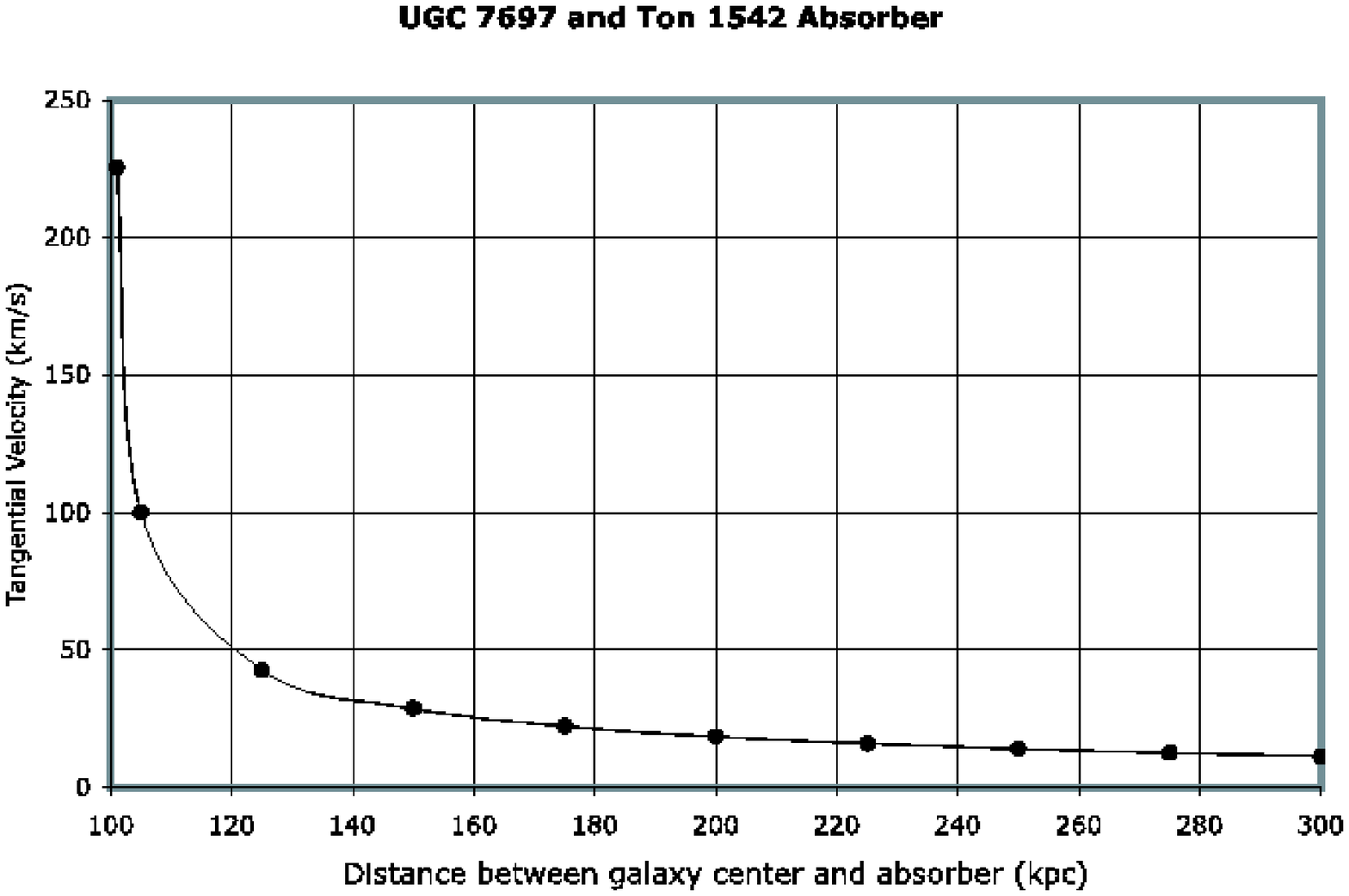}
\includegraphics[angle=0,scale=.6]{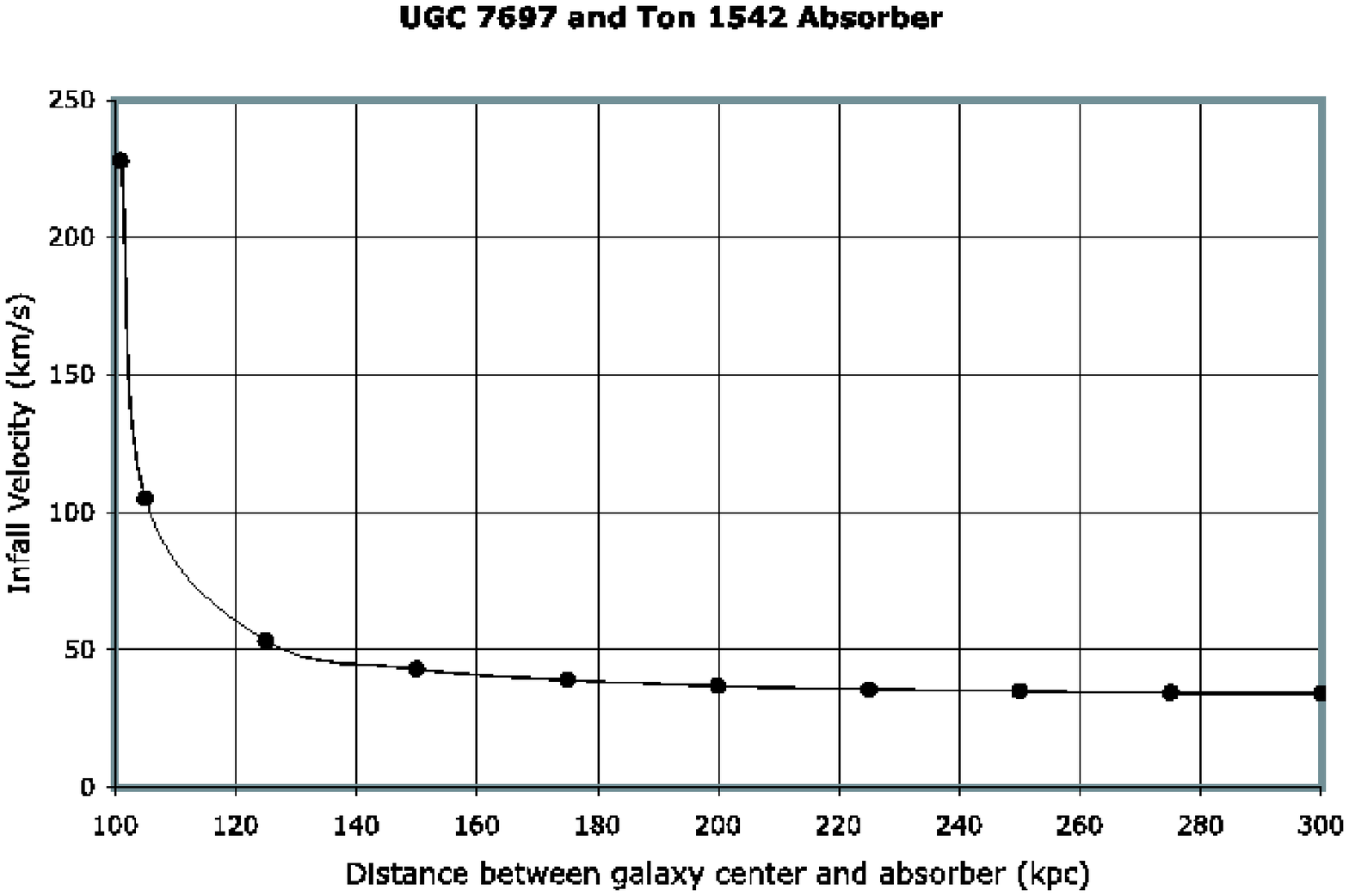}
\caption{The top panel shows the magnitude of the minimum tangential velocity at each distance from 
the center of UGC 7697
for the gas traced by the absorber Ton 1542 to be infalling (only if the gas is in front of the galaxy) or
outflowing (only if the gas is behind the galaxy) and no longer counter rotating.  The bottom panel
shows the magnitude of the corresponding minimum 
infall/outflow velocity.  The minimum magnitude of the V$_{\rm los}$ component
of the absorber relative to UGC 7697 is 32 \kms.}
\end{figure}

\clearpage


\clearpage

\begin{figure}
\includegraphics[angle=0,scale=.6]{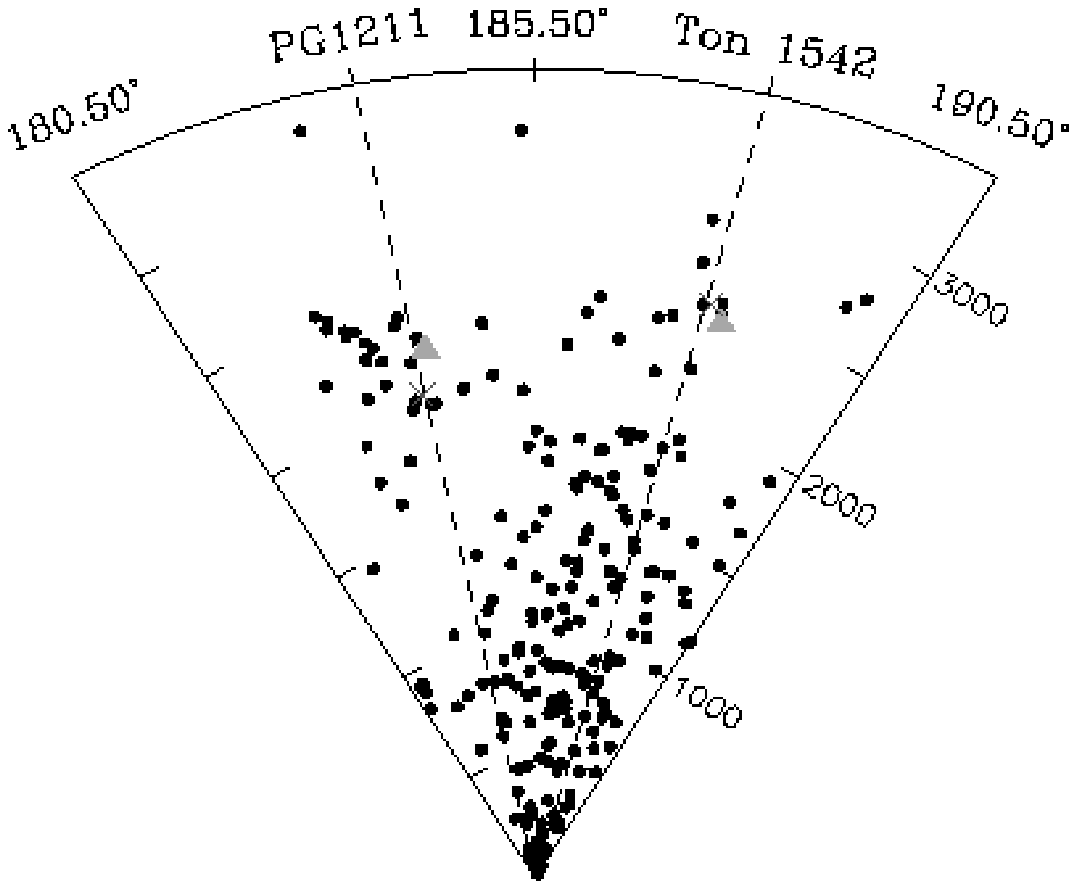}
\includegraphics[angle=0,scale=.6]{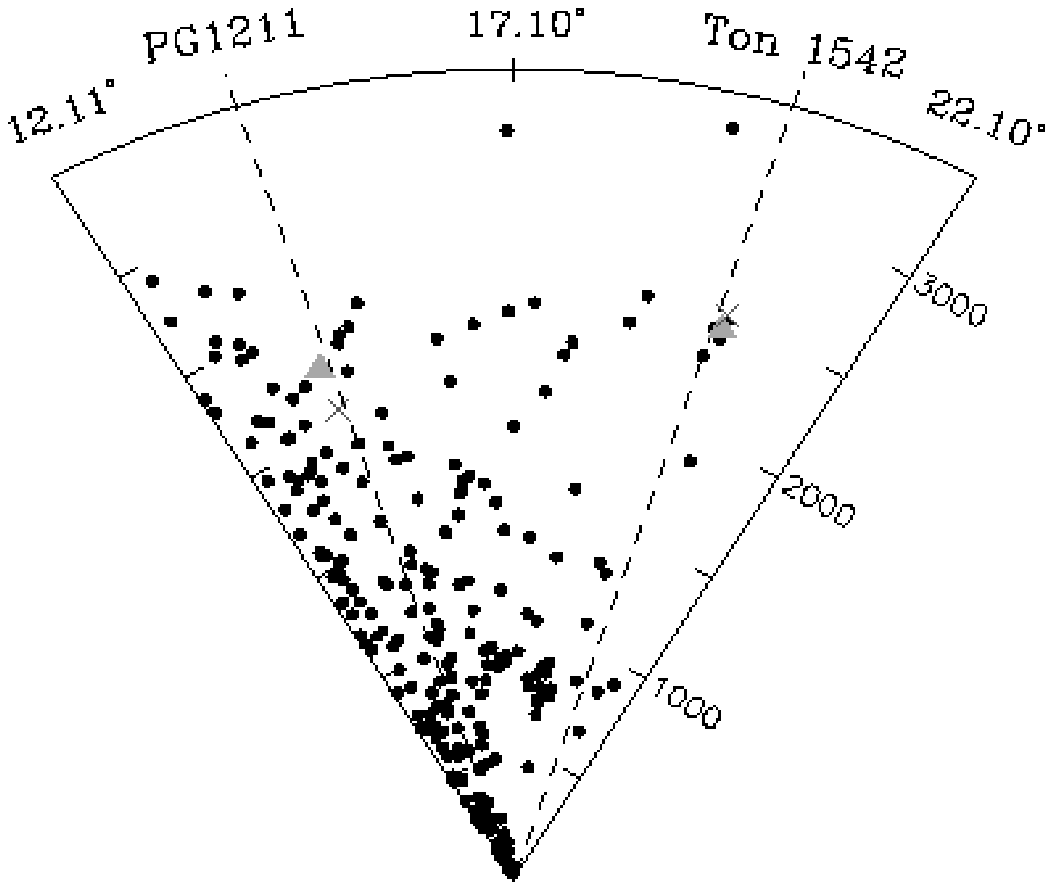}
\caption{The cone diagrams showing the local large scale structure for both the PG 1211+143 
and Ton 1542 absorber sightlines for a 3500 \kms~velocity
range around the galaxies of interest here (IC 3061 and UGC 7697 respectively).
The filled circles are galaxies from the Updated Zwicky Catalog 
(Falco et al. 1999), the $\times$s represent the positions of the absorbers, and the triangles represent the positions of IC 3061 
and UGC 7697. The top panel shows velocity along the radial direction and right ascension along the arcs. 
The plot is 10 degrees deep in declination centered between the Ly$\alpha$ absorbers.  The bottom panel shows 
declination versus velocity, in this case for a 10 degree wide strip in right ascension 
centered between the absorbers.}
\end{figure}

\end{document}